\documentclass[a4paper,twoside,twocolumn,showpacs,superscriptaddress]{revtex4}
\usepackage{color}
\usepackage{amsmath}
\usepackage{amssymb}
\usepackage{graphicx}
\usepackage{hyperref}

\begin{document}

\title{Analytical and numerical study of dirty bosons in a\\ quasi-one-dimensional harmonic trap}

\author{Tama Khellil}
\email{khellil.lpth@gmail.com}
\affiliation{Institute for Theoretical Physics, Free University of Berlin, Arnimallee 14, 14195 Berlin, Germany}

\author{Antun Bala\v{z}}
\email{antun.balaz@ipb.ac.rs}
\affiliation{Scientific Computing Laboratory, Institute of Physics Belgrade, University of Belgrade, Pregrevica 118, 11080 Belgrade, Serbia}

\author{Axel Pelster}
\email{axel.pelster@physik.uni-kl.de}
\affiliation{Department of Physics and Research Center OPTIMAS, Technical University of Kaiserslautern, 67663 Kaiserslautern, Germany}

\begin{abstract}
The emergence of a Bose-glass region in a quasi one-dimensional Bose-Einstein-condensed
gas in a harmonic trapping potential with an additional delta-correlated
disorder potential at zero temperature is studied using three approaches.
At first, the corresponding time-independent Gross-Pitaevskii equation
is numerically solved for the condensate wave function, and disorder
ensemble averages are evaluated. In particular, we analyse quantitatively
the emergence of mini-condensates in the local minima of the random
potential, which occurs for weak disorder preferentially at the border
of the condensate, while for intermediate disorder strength this happens
in the trap centre. Second, in view of a more detailed physical understanding
of this phenomenon, we extend a quite recent non-perturbative approach
towards the weakly interacting dirty boson problem, which relies on
the Hartree-Fock theory and is worked out on the basis of the replica
method, from the homogeneous case to a harmonic confinement. Finally,
in the weak disorder regime we also apply the Thomas-Fermi approximation,
while in the intermediate disorder regime we additionally use a variational
ansatz in order to describe analytically the numerically observed
redistribution of the fragmented mini-condensates with increasing
disorder strength. 
\end{abstract}


\maketitle

\section{Introduction}
\label{I}

The dirty boson problem is defined as a system of interacting bosons
in a random potential \cite{Intro-11}. The combined effect of disorder
and two-particle interaction represents one of the most challenging
problems in condensed matter physics due to the intriguing interplay
between localisation and superfluidity. Cold atoms provide a controlled
experimental setup in which that fundamental question of interacting
bosons in a random environment can be addressed in both a quantitative
and a tunable way.

The earliest relevant experiments, which were central for motivating
the research of the dirty boson problem, dealt with superfluidity
of thin films of $^{4}$He adsorbed in porous Vycor glass in the low-density
limit \cite{Intro-91}. There it was proven that, despite the presence
of disorder, superfluidity can still persist. For ultracold Bose gases
disorder appears either naturally as, e.g., in magnetic wire traps
\cite{Intro-75,Intro-76,Intro-77,Fortagh,Schmiedmayer}, where imperfections
of the wire itself can induce local disorder, or it may be created
artificially and controllably as, e.g., by using laser speckle fields
\cite{Intro-78,Intro-79,Intro-21,Goodmann,Intro-16}. A set-up more
in the spirit of condensed matter physics relies on a Bose gas with
impurity atoms of another species trapped in a deep optical lattice,
so the latter represent randomly distributed scatterers \cite{Intro-13,Intro-14}.
Furthermore, an incommensurate optical lattice can provide a pseudo-random
potential for an ultracold Bose gas \cite{Lewenstein,Ertmer,intro-17}.

Non-interacting particles in a random environment can be localised
provided that the disorder is sufficiently strong. This phenomenon
of Anderson localisation occurs as the particles are repeatedly reflected
back in the random potential, so interferences yield exponentially
localised one-body wave functions \cite{Intro-15}. In one dimension
Anderson localisation was experimentally found in an ultracold Bose
gas in references \cite{Intro-16,intro-17}. Within a Bose-Einstein condensation
(BEC), which is a many-particle interacting system, the presence of
disorder causes the emergence of a new phase besides the superfluid
phase (SF), which is called a Bose-glass phase due to the localisation
of bosons in the respective minima of the random potential landscape.
This Bose-glass phase contains no superfluid fraction and is characterised
by a finite compressibility, by the absence of a gap, and by an infinite
superfluid susceptibility \cite{Intro-11}. Indications for the existence
of the Bose-glass phase were found, for instance, in the experiments
of references \cite{Fortagh,Schmiedmayer,Intro-79,Ref2-2}. There
it was shown within the superfluid phase that an increasing disorder
strength yields first a fragmentation of the condensate due to the
formation of tiny BEC droplets in the minima of the random environment.
For sufficiently strong disorder the condensate then turns out to
be completely destroyed as all bosons are localised in the minima
of the random potential, which represents the Bose-glass phase. But
a more quantitative investigation of that elusive phase is still lacking
both from an experimental and a theoretical point of view.

One of the first important theoretical results of the dirty boson
problem was obtained by Huang and Meng in 1992 \cite{HM-1}. Within
a Bogoliubov theory for a weakly interacting Bose-Einstein condensate
it was found that a weak random disorder potential leads to a depletion
of both the global condensate density and the superfluid density due
to the localisation of bosons in the respective minima of the random
potential \cite{HM-1,HM-3,HM-4,HM-5,HM-2,HM-8,HM-6,HM-9,HM-10,HM-7,Ref2-3,Ref2-7}.
Beyond the weak disorder, a perturbative approach was worked out in
references \cite{Gaul-Muller-1,Gaul-Muller-2}, where the impact of the
external random potential upon the quantum fluctuations was studied
in detail. In order to analyse the BEC in the strong disorder regime,
there are, in principle, two complementary non-perturbative approaches.
The first starts from the superfluid phase and ends up in the Bose-glass
phase for increasing disorder strength. To this end reference \cite{Navez}
applies the random phase approximation and yields a self-consistent
integral equation for the disorder-averaged particle density, whereas
references \cite{Yukalov1,Yukalove2} work out a stochastic self-consistent
mean-field approach using two chemical potentials, one for the condensed
and one for the exited particles. The second approach starts conversely
from the Bose-glass phase and proceeds towards the superfluid phase
for decreasing disorder strength. For instance, references \cite{Nattermann,Natterman2}
work this out on the basis of a careful energetic analysis, where
the disorder turns out to strongly influence the size, shape, and
structure of the mini-condensates in the minima of the random potential.
Furthermore, an order parameter was introduced and applied in the
context of a Hartree-Fock theory in reference \cite{Intro-90} in order
to describe the possible emergence of the Bose-glass phase.

The localisation due the presence of the disorder in one-dimensional BEC systems was tackled in the
literature in different directions. For instance, it was shown analytically
that the one-dimensional gas of short-range-interacting bosons in
the presence of disorder can undergo a finite-temperature phase transition
between superfluid and insulator \cite{Ref2-1}. Furthermore, solving numerically
and variationally the Gross-Pitaevskii equation of the weakly interacting
BEC in a weakly disordered lattice and a speckle potential, the localised
BEC’s are found to have an exponential tail \cite{Ref1-4}. Using
quantum Monte Carlo simulations it turned out that, surprisingly,
disorder-induced phase coherence could occur \cite{Ref1-1}. Furthermore,
in the context of optical lattices, the quantum phase diagram of a
dirty BEC in one dimension was also investigated via different methods
\cite{Intro-11}. Reference~\cite{Ref1-2} proved that approximative
description of all quantum phases can be obtained via the site-dependent
decoupling mean-field approach. By means of the density matrix renormalisation
group technique, the existence of a critical value of the disorder
strength for the Bose-glass phase was proven in reference \cite{Ref1-3}.
In addition, the exact Bose-Fermi mapping demonstrated that the superfluid
Bose-glass transition and the general phase diagram of trapped incommensurate
optical lattices can be uniquely determined from finite-temperature
density distributions of the trapped disordered system \cite{Ref1-5}.
Despite all those previous investigations there is still a lack of
knowledge concerning the emergence of the Bose-glass phase and its
elusive properties.

In the present paper we treat a quasi-one-dimensional trapped BEC in
a disorder potential both analytically and numerically.
In particular, we focus on the question how the bosons, which are localised in the minima of the random potential,
are distributed within the harmonic confinement. For sufficiently large disorder we even expect to find a Bose-glass
region in the trap, where the global condensate vanishes and only localised bosons exist.
Note that the corresponding three-dimensional trapped case is treated separately in reference \cite{Paper3}.
We start first by describing the underlying BEC model and by developing
a Hartree-Fock mean-field theory for the weak disorder regime and
apply it within the Thomas-Fermi approximation to the dirty BEC system
in section \ref{II}. However, since we study a system that is not fully amenable to the Thomas-Fermi approximation,
we also employ numerical and variational treatment, described
in section \ref{III}. We solve the corresponding Gross-Pitaevskii
(GP) equation of the BEC model, and then apply a variational ansatz
for the intermediate disorder regime. The results of those three different
methods are discussed and compared in section \ref{IV}.
For instance, we find that the density of fragmented mini-condensates is redistributed for increasing disorder strength.
Whereas for small disorder bosons tend to localise at the border of the trap, for intermediate disorder strength they
concentrate in the trap centre.

\section{Hartree-Fock mean-field theory in 1D}
\label{II}

It has been suggested in reference \cite{HM-4} that a Gaussian-correlated
disorder potential constitutes an appropriate model to describe realistic random landscapes.
Therein the final correlation length corresponds to the average width of the mountains or valleys in the disorder potential.
Such a Gaussian correlation has been explored in more detail both for a BEC with contact as well as dipole-dipole
interaction in any geometry \cite{HM-7,HM-8,Ref2-3}. Qualitatively similar results
are obtained for other disorder potentials with finite correlation lengths as, for instance,
laser speckles \cite{HM-6,HM-7} and Lorentz correlation \cite{HM-9}. All these studies
have in common that disorder effects typically decrease with increasing correlation length and are, thus,
most pronounced for $\delta$-correlation. Therefore, we restrict ourselves in the following to the case
of $\delta$-correlation, to focus on the study of disorder effects.

The model of a three-dimensional weakly interacting homogeneous Bose
gas in a $\delta$-correlated disorder potential was studied within
the Hartree-Fock mean-field theory in reference \cite{Intro-90} by applying
the Parisi replica method \cite{Intro-33,Intro-36,Intro-37}. As a
result, the corresponding phase diagram for the occurrence of the
superfluid, the Bose-glass, and the normal phase was determined in
the control parameter plane spanned by disorder strength and temperature.
This Hartree-Fock theory is extended in reference \cite{Paper2} to a
harmonic confinement, and is applied in the following to one-dimensional systems.

To this end, we consider a model of one-dimensional harmonically trapped BEC in
a $\delta$-correlated disorder potential with contact interactions
between the particles. The corresponding free energy is calculated
in Appendix~\ref{F} and is given by equation \eqref{F1}. It depends on the
superfluid order parameter $n_0(x)$, which represents the condensate density,
the Bose-glass order parameter $q(x)$, which stands for the density of atoms being localised
in the local minima of disorder potential, and an auxiliary function $Q_0(x)$. The self-consistency equations
are obtained by extremizing the free energy with respect to these
functions, i.e., $\frac{\delta\mathcal{F}}{\delta n_0(x')}=0$,
$\frac{\delta\mathcal{F}}{\delta Q_0(x')}=0$, and $\frac{\delta\mathcal{F}}{\delta q(x')}=0$.
This yields, together with the particle number equation \eqref{3.5}, four coupled equations:
an algebraic equation for $q(x)$,
\begin{equation}
q(x)=\frac{D}{\hbar M}Q_0(x)^3\,\frac{n_0(x)}{1-\frac{D}{\hbar M}Q_0(x)^3}\, ,\label{q1d}
\end{equation}
a nonlinear differential equation for $n_0(x)$, 
\begin{eqnarray}
\Biggl[-\mu+2gn(x)+V(x)-gn_0(x)-\frac{D}{\hbar}Q_0(x)\nonumber \\
-\frac{\hbar^2}{2M}\frac{\partial^2}{\partial x^2}\Biggr]\,\sqrt{n_0(x)}=0\, ,\label{PMM-1}
\end{eqnarray}
the equation for the total density $n(x)$, which is sum of the previous
two densities, 
\begin{eqnarray}
n(x)=q(x)+n_0(x)\, ,\label{MAT1-1}
\end{eqnarray}
and the auxiliary function $Q_0(x)$, which is a solution of the
cubic equation 
\begin{equation}
-\frac{D}{\hbar}Q_0(x)^3+\left[-\mu+2gn(x)+V(x)\right]Q_0(x)^2-\frac{M}{2}=0\, .\label{Q3}
\end{equation}
Here $D$ denotes the strength of disorder, as defined in Appendix~\ref{F}.
Note that in the clean case ($D=0$) equations \eqref{q1d}--\eqref{Q3} reduce to the standard
Gross-Pitaevskii theory. Furthermore, for finite disorder strength $D$ the homogeneous case,
where $V(x)=0$, is treated semi-analytically in Appendix~\ref{homogeneous}. There it is shown that
increasing the disorder strength $D$ yields a first-order quantum phase transition from the superfluid to the
Bose-glass phase.

For the harmonically trapped case, however, no analytic approach is known which gives the exact
solution of the differential equation \eqref{PMM-1} even in the
absence of disorder. Therefore, we approximate its solution via the
Thomas-Fermi (TF) approximation method, which is based on neglecting
the kinetic energy. To this end, it turns out that we have to distinguish between two different spatial
regions: the superfluid region, where the bosons are distributed in
the condensate as well as in the minima of the disorder potential,
and the Bose-glass region, where there are no bosons in the global
condensate so that all bosons contribute only to the local Bose-Einstein
condensates. In the following the radius of the superfluid region,
i.e., the condensate radius, is denoted by $R_{\rm TF1}$, while
the radius of the whole bosonic cloud $R_{\rm TF2}$ is called
the cloud radius.

Within the TF approximation the algebraic equations \eqref{q1d},
\eqref{MAT1-1}, and \eqref{Q3} remain the same, but the differential
equation \eqref{PMM-1} reduces to an algebraic relation in the superfluid
region: 
\begin{eqnarray}
-\mu+2gn(x)+V(x)-gn_0(x)-\frac{D}{\hbar}Q_0(x)=0\, .\label{PMM-1-1-1}
\end{eqnarray}
Outside the superfluid region, i.e., in the Bose-glass region, equation
\eqref{PMM-1} reduces simply to $n_0(x)=0$. The advantage of
the TF approximation is that now we have only four coupled algebraic
equations.

At first we consider the superfluid region. In the TF approximation
the dependency on the auxiliary function $Q_0(x)$ in equation \eqref{Q3}
can be eliminated and equations \eqref{q1d}, \eqref{MAT1-1}, and \eqref{PMM-1-1-1}
reduce in the superfluid region to: 
\begin{eqnarray}
&&-\tilde{\mu}+2\tilde{n}(\tilde{x})+\tilde{x}^2-\tilde{n}_0(\tilde{x})-2\frac{\tilde{D}}{\sqrt{\tilde{n}_0(\tilde{x})}}=0\, ,\label{PMM-1-1-1-1}\\
&&\tilde{q}(\tilde{x})=\tilde{D}\,\frac{\tilde{n}_0(\tilde{x}}{\tilde{n}_0(\tilde{x})^{3/2}-\tilde{D}}\, ,\label{q1d-1-1}\\
&&\tilde{n}(\tilde{x})=\tilde{q}(\tilde{x})+\tilde{n}_0(\tilde{x})\, ,\label{MAT1-1-1-1}
\end{eqnarray}
where $\tilde{n}(\tilde{x})=n(x)/\overline{n}$ denotes the dimensionless
total density, $\tilde{n}_0(\tilde{x})=n_0(x)/\overline{n}$ the
dimensionless condensate density, $\tilde{q}(\tilde{x})=q(x)/\overline{n}$
the dimensionless Bose-glass order parameter, $\tilde{x}=x/R_{{\rm {TF}}}$
the dimensionless coordinate, $\overline{n}=\bar{\mu}/g$ the maximal
total density in the clean case, $\tilde{\mu}=\mu/\bar{\mu}$ the
dimensionless chemical potential, $\tilde{D}=\frac{\xi^{3}}{\mathcal{L}^{3}}$
the dimensionless disorder strength, $\xi=\frac{l^2}{R_{{\rm {TF}}}}$
the coherence length in the centre of the trap, $l=\sqrt{\frac{\hbar}{M\Omega}}$
the oscillator length, and $R_{{\rm {TF}}}=l\sqrt{\frac{2\bar{\mu}}{\hbar\Omega}}$
the TF cloud radius in the clean case, i.e., when
$D=0$. The chemical potential in the absence of the disorder $\bar{\mu}=\hbar\omega_{r}\left(\frac{3}{2\sqrt{2}}N\frac{a}{l}\sqrt{\frac{\Omega}{\omega_{r}}}\right)^{2/3}$,
which provides the energy scale, is deduced from the normalisation
condition \eqref{3.5} in the clean case.

Inserting equations \eqref{q1d-1-1} and \eqref{MAT1-1-1-1} into equation \eqref{PMM-1-1-1-1}
gives us one self-consistency equation for the condensate density
in the superfluid region: 
\begin{eqnarray}
\tilde{n}_0(\tilde{x})^3+\left(-\tilde{\mu}+\tilde{x}^2\right)\tilde{n}_0(\tilde{x})^2-\tilde{D}\tilde{n}_0(\tilde{x})^{3/2}\nonumber \\
-\tilde{D}\left(-\tilde{\mu}+\tilde{x}^2\right)\sqrt{\tilde{n}_0(\tilde{x})}+2\tilde{D}^2=0\, .\label{PMM-1-1-1-1-1}
\end{eqnarray}

This equation is of sixth order with respect to $\sqrt{\tilde{n}_0(\tilde{x})}$,
which makes it impossible to solve analytically. Therefore, we solve
it numerically and insert the result into equations \eqref{q1d-1-1} and
\eqref{MAT1-1-1-1} in order to determine the Bose-glass order parameter
$\tilde{q}(\tilde{x})$ and the total density $\tilde{n}(\tilde{x})$,
respectively.

Now we come to the Bose-glass region, where we have from equation \eqref{PMM-1}
in TF approximation $\tilde{n}_0(\tilde{x})=0$, and equation \eqref{MAT1-1}
reduces to $\tilde{n}(\tilde{x})=\tilde{q}(\tilde{x})$. Inserting
this into equation \eqref{q1d}, we get $Q_0(x)$=$\left(\frac{\hbar M}{D}\right)^{1/3}$,
which reduces equation \eqref{Q3} to: 
\begin{equation}
\tilde{q}(\tilde{x})=\frac{1}{2}\left(3\tilde{D}^{2/3}+\tilde{\mu}-\tilde{x}^2\right)\, .\label{bg}
\end{equation}

We also need to write down the dimensionless equivalent of the normalisation
condition $\eqref{3.5}$, which reads: 
\begin{equation}
\int_{-\tilde{R}_{\rm TF2}}^{\tilde{R}_{\rm TF2}}\tilde{n}(\tilde{x})d\tilde{x}=\frac{4}{3}\, ,\label{18-1}
\end{equation}
where $\tilde{R}_{\rm TF2}=R_{\rm TF2}/R_{\rm TF}$ denotes
the dimensionless cloud radius, and the total density $\tilde{n}(\tilde{x})$
in equation \eqref{18-1} is the combination of the total densities from
both the superfluid region and the Bose-glass region.

Before considering any particular parameter value for our BEC system,
we have first to justify using the TF approximation and determine
its range of validity. To this end we rewrite equation \eqref{PMM-1} in
the clean case, i.e., for $D=0$, and divide it with $\bar{\mu}$.
This yields:

\begin{equation}
\left[-1+\tilde{n}(\tilde{x})+\tilde{x}^2-\left(\frac{\xi}{R_{\rm TF}}\right)^2\frac{\partial^2}{\partial\tilde{x}^2}\right]\,\sqrt{\tilde{n}(\tilde{x})}=0\, .\label{PMM-1-1-3}
\end{equation}
Note that in the clean case the total density coincides with the condensate one.
The TF approximation is only justified when the prefactor of
the kinetic term $\left(\frac{\xi}{R_{\rm TF}}\right)^2$ is
small enough, so the kinetic term can be neglected, which yields 
\begin{equation}
\xi\ll R_{\rm TF}\, .\label{TF}
\end{equation}
The corresponding TF results are presented and discussed
in section \ref{IV}. In order to asses their validity, in particular for large disorder strengths, however,
we turn first to two other complementary approaches to the dirty BEC problem.

\section{Numerical and variational approach}
\label{III}

In this section we start with working out a numerical method that relies on solving the Gross-Pitaevskii
equation for an ensemble of realisations of disorder landscape. Furthermore, we also introduce a variational approach,
which is tailored to describe the numerical results analytically.

\subsection{Numerical method}

Now we perform a numerical study for the Bose-condensed gas in one
dimension at zero temperature in a harmonic trapping potential $V(x)=\frac{1}{2}M\Omega^2x^2$.
Furthermore, we assume a Gaussian-distributed disorder potential $U(x)$,
which satisfies the conditions
\begin{equation}
\overline{U(x)}=0\, ,\label{eq:19}
\end{equation}
and 
\begin{equation}
\overline{U(x)U(x')}=D(x-x')\, ,\label{1}
\end{equation}
where $D(x-x')$ denotes the correlation function.

A one-dimensional BEC in the mean-field Hartree approximation is given
by a generalised time-independent Gross-Pitaevskii equation for the
condensate wave function $\psi(x)$: 
\begin{equation}
\left[-\frac{\hbar^2}{2M}\frac{\partial^2}{\partial x^2}-\mu+U(x)+V(x)+g|\psi(x)|^2\right]\psi(x)=0\, .\label{20}
\end{equation}

Equation \eqref{20} represents a stochastic nonlinear
differential equation which can not be solved exactly, and, therefore,
we apply a numerical approach. To this end we have first to generate
the random potential $U(x)$ before inserting it into equation $\left(\ref{20}\right)$,
and then calculate the disorder average over many realisations of
$U(x)$.

Motivated by Fourier series, a simple ansatz for generating a random
Gaussian function $U(x)$ is performed as follows.
The potential is written as a finite superposition of $\sin kx$
and $\cos kx$ terms with properly selected amplitudes $A_n$,
$B_n$, and wave numbers $k_n$ \cite{Generating,Duttman}:

\begin{equation}
U(x)=\frac{1}{\sqrt{\mathtt{N}}}\sum_{n=0}^{\mathtt{N}-1}\left(A_n\cos k_n x+B_n \sin k_n x\right),\label{8}
\end{equation}
where $\mathbb{\mathtt{N}}$ denotes the number of terms, which should
be large enough in order to obtain a good approximation for the random
potential. Furthermore, we assume$\ A_{n}$ and$\ B_{n}\ $to be mutually
independent Gaussian random variables with zero mean, and variance
equal to $D(0)$: 
\begin{equation}
\overline{A_n B_n}=0\, , \qquad \overline{A_n A_m}=\overline{B_n B_m}=D(0)\delta_{nm}\, .\label{A-B}
\end{equation}

\begin{figure*}[!ht]
\includegraphics[width=0.65\textwidth]{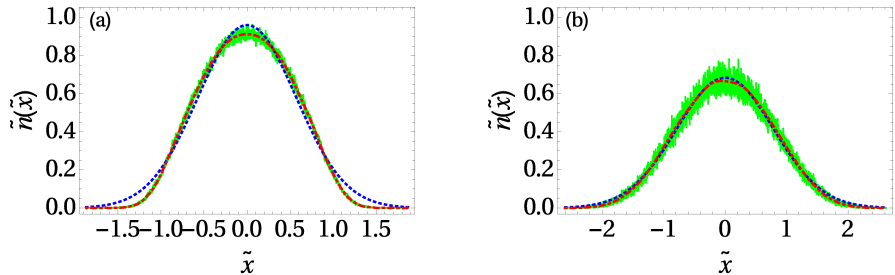}
\caption{Spatial distribution of the particle density
$\tilde{n}(\tilde{x})$: numerical data (solid, green), fitted curve
(dotted-dashed, red), and fitted Gaussian (dotted, blue) for (a) $\tilde{D}=0.067$
and (b) $\tilde{D}=0.603$.
\label{fig:1}}
\end{figure*}

The wave numbers $k_n$ are independent random variables, as well,
selected from the probability distribution: 
\begin{equation}
p(k_n)=\frac{S(k_n)}{\int_{-\infty}^\infty S(k')dk'}\, ,\label{p}
\end{equation}
where $S(k)$ defines the spectral density as the Fourier transform
of the correlation function:

\begin{equation}
S(k)=\int_{-\infty}^\infty dx\, e^{-ikx}D(x)\, .\label{S}
\end{equation}
In the special case of the Gaussian-correlated disorder we have 
\begin{equation}
D(x-x')=\frac{D}{\sqrt{2\pi}\lambda}\, e^{-\frac{(x-x')^2}{2\lambda^2}}\, ,\label{1-2}
\end{equation}
where $\lambda$ denotes the correlation length and $D$ the disorder
strength. The probability distribution \eqref{p} reads in this case:
\begin{equation}
p(k_n)=\frac{\lambda}{\sqrt{2\pi}}\, e^{-\frac{\lambda^2 k_n^2}{2}}\, .\label{p-1}
\end{equation}
Note that the analytical study in section \ref{II} is done for $\delta$-correlated
disorder, but since it is impossible to treat the $\delta$-correlated
disorder numerically, we use the Gaussian-distributed disorder \eqref{1-2},
which specialises to a $\delta$-distributed one in the limit $\lambda\to 0$,
i.e., $\lim_{\lambda\to 0}D\left(x\right)=D\delta(x).$

In order to numerically generate the correlation function \eqref{1-2}
with sufficient accuracy, two numbers have to be appropriately large
enough. The first one is the number $\mathtt{N}$ of terms in equation
\eqref{8}, the second one is the number $\mathtt{M}$ of realisations
of the disorder potential, which are used to evaluate the disorder
ensemble average \eqref{1-2}. It can be shown analytically that the
error in reproducing the correlation function \eqref{1-2} in the
case $\mathtt{M}\to\infty$ is of the order of $\mathtt{1/N}$
\cite{Duttman}. All Gaussian-correlated quantities are generated
using the Box-M\" ulller algorithm \cite{Box-Mueller}. We
insert the generated disorder potential \eqref{8} into the Gross-Pitaevskii
equation \eqref{20}, and then use a C computer program that solves the
time-independent Gross-Pitaevskii equation in one space dimension
in a harmonic trap using the imaginary-time propagation \cite{Fort,GP-SCL,GP-DBEC,GP-DBEC-CUDA,GP-SCL-HYB}.
In this way we obtain the numerical solution of the ground-state wave
function $\psi(x)$ of equation \eqref{20} for $\mathtt{M}=1000$ realisations
of the disorder potential and $\mathtt{N}=10000$ terms in equation \eqref{8}.
To this end we use different values of the disorder strength $D$
in order to cover the range from the weak to the intermediate disorder
regime. We have chosen the disorder correlation length to be $\lambda=0.01\, l$,
which is small enough in order to approach the case of $\delta$-correlated
disorder.

Performing disorder ensemble averages, we have access to the particle
density $n(x)=\overline{\psi(x)^2}$, to the condensate density
$n_0(x)=\overline{\psi(x)}^2$, and to the Bose-glass order
parameter $q(x)=n(x)-n_0(x)$. In order to compare the
numerical results with the analytical ones obtained in section \ref{II},
we use the same rescaling parameters for all densities, coordinates,
chemical potential, and disorder strength, as already explained
below equation \eqref{MAT1-1-1-1}.

Before discussing the numerical results in detail, we show first one
typical example in two graphs in figure~\ref{fig:1}, where the total density $\tilde{n}(\tilde{x})$
is plotted for two different values of the disorder strength (solid,
green line), showing the original data for $\mathtt{M}=1000$ and
$\mathtt{N}=10000$ terms in equation \eqref{8}. We remark that the resulting
density is fluctuating around a Gaussian-like curve. Comparing figure~\ref{fig:1}(a)
with figure~\ref{fig:1}(b) we conclude that the fluctuations are increasing
with the disorder strength. The origin of those fluctuations is that
the $\mathtt{M}=1000$ realisations of the disorder potential for
performing the disorder ensemble average are not sufficient to produce
a smooth curve. One solution of this problem would be to increase
the number $\mathtt{M}$ of the realisations of the disorder potential,
which would need longer execution time, especially for the intermediate
disorder regime, where the numerics has to be run for a larger spatial
range. Another solution is to extract a continuous smooth curve that
fits best to our data, as it is done in figure~\ref{fig:1} (dotted-dashed,
red line). This method is applied to all numerical densities in this
paper. Furthermore, from the Gaussian fit in figure~\ref{fig:1} (dotted,
blue line), we remark that the original data of the total density
approach a Gaussian form in the intermediate disorder regime much
better than in the weak disorder regime. This can be explained with
the argument that increasing the disorder reduces effectively the
repulsive interaction between the particles \cite{Paper2} and, thus, approaches
the case of non-interacting bosons, where the total density is given
by a Gaussian.

\subsection{Variational method}

Since the four self-consistency equations \eqref{q1d}--\eqref{Q3} are
obtained by extremizing the free energy \eqref{F1}, we can apply
the variational method in the spirit of references \cite{Kleinert1,Kleinert2,Var1,Var2,Alex1,Alex2,Alex3,Alex4}
to obtain approximate results. In order to be able to compare the
variational results with the analytical and the numerical ones from
section \ref{II} and the previous subsection, respectively, we use
the same rescaling parameters already introduced below equation \eqref{MAT1-1-1-1}
for all functions and parameters. To this end, we have to multiply
\eqref{F1} with the factor $1/(\bar{\mu}\overline{n}R_{\rm TF})$
to obtain: 
\begin{widetext}
\begin{eqnarray}
\tilde{\mathcal{F}} & = & \int d\tilde{x}\Bigg[-[\tilde{q}(\tilde{x})+\tilde{n}_0(\tilde{x})]^2-\sqrt{\tilde{n}_0(\tilde{x})}
\left\{\tilde{\mu}+\left(\frac{\xi}{R_{\rm TF}}\right)^2\frac{\partial^2}{\partial\tilde{x}^2}-2[\tilde{q}(\tilde{x})+\tilde{n}_0(\tilde{x})]-\tilde{x}^2
 +2\tilde{D}\tilde{Q}_0(\tilde{x})\right\}\sqrt{\tilde{n}_0(\tilde{x})}\nonumber\\
 &&-\frac{1}{2}\tilde{n}_0(\tilde{x})^2+2\tilde{D}\tilde{Q}_0(\tilde{x})[\tilde{q}(\tilde{x})+\tilde{n}_0(\tilde{x})]
 -2\tilde{D}\frac{\tilde{q}(\tilde{x})+\tilde{n}_0(\tilde{x})}{\sqrt{-\tilde{\mu}+2[\tilde{q}(\tilde{x})+\tilde{n}_0(\tilde{x})]+\tilde{x}^2-2\tilde{D}\tilde{Q}_0(\tilde{x})}}\Bigg] ,\label{F1-3}
\end{eqnarray}
\end{widetext}
where $\tilde{\mathcal{F}}=\mathcal{F}/(\bar{\mu}\bar{n}R_{\rm TF})$
denotes the dimensionless free energy and $\tilde{Q}_0(\tilde{x})=\sqrt{\frac{2\bar{\mu}}{M}}Q_0(x)$.

Motivated by the numerical results presented in figure~\ref{fig:1}, we
suggest the three following ans\" atze for the condensate density $\tilde{n}_0\left(\tilde{x}\right)$,
the Bose-glass order parameter $\tilde{q}\left(\tilde{x}\right)$,
and the auxiliary function $\tilde{Q}_0(\tilde{x})$: 
\begin{eqnarray}
\tilde{n}_0\left(\tilde{x}\right)&=&\alpha e^{-\sigma\tilde{x}^2}\, ,\label{n0}\\
\tilde{q}\left(\tilde{x}\right)+\tilde{n}_0\left(\tilde{x}\right)&=&\gamma e^{-\theta\tilde{x}^2}\, ,\label{q-1}\\
\tilde{Q}_0(\tilde{x})&=&\frac{\tilde{q}\left(\tilde{x}\right)+\tilde{n}_0\left(\tilde{x}\right)}{\tilde{D}}-\left(\zeta+\eta\tilde{x}^2\right)\, ,\label{Q0-2}
\end{eqnarray}
where $\alpha$, $\sigma$, $\gamma$, $\theta$, $\zeta$, and $\eta$
denote variational parameters. The parameters $\alpha$ and $\gamma$
are proportional to the number of particles in the condensate and
the total number of particles, while parameters $\sigma$ and $\theta$
represent the width of the condensate density and the total density,
respectively.

Inserting the ans\" atze \eqref{n0}--\eqref{Q0-2} into the free energy
\eqref{F1-3} and performing the integral yields:
\begin{eqnarray}
&&\tilde{\mathcal{F}}  = \sqrt{\pi}\Bigg\{ \frac{\gamma^2}{\sqrt{2\theta}}+\frac{\alpha}{2\sigma^{3/2}}-\frac{\alpha}{4\sqrt{\sigma}}\left(4\tilde{\mu}+\sqrt{2}\alpha\right)\nonumber \\
 &  & +\left(\frac{\xi}{R_{\rm TF}}\right)^2\frac{\alpha\sqrt{\sigma}}{2}+\tilde{D}\left(\frac{2\alpha\zeta}{\sqrt{\sigma}}+\frac{\alpha\eta}{\sigma^{3/2}}-\frac{\gamma(\eta+2\zeta\theta)}{\theta^{3/2}}\right)\nonumber \\
 &  & -\frac{2\tilde{D}\gamma}{\sqrt{1+2\tilde{D}\eta}}e^{\frac{2\tilde{D}\zeta-\tilde{\mu}}{2+4\tilde{D}\eta}\theta}K_0\left(\frac{2\tilde{D}\zeta-\tilde{\mu}}{2+4\tilde{D}\eta}\theta\right)\Bigg\} ,\label{F1-4}
\end{eqnarray}
where $K_0(s)$ represents the modified Bessel function of the second kind.

The free energy \eqref{F1-4} has now to be extremized with respect
to the variational parameters $\alpha$, $\sigma$, $\gamma$, $\theta$,
$\zeta$, and $\eta$. Together with the thermodynamic condition $-\frac{\partial\tilde{\mathcal{F}}}{\partial\tilde{\mu}}=\frac{4}{3}$,
we have seven coupled equations for seven variables $\alpha$, $\sigma$,
$\gamma$, $\theta$, $\zeta$, $\eta$, and $\tilde{\mu}$ that we
solve numerically. From all physical solutions we select the one with
the smallest free energy \eqref{F1-4}, then we insert the resulting
variational parameters $\alpha$, $\sigma$, $\gamma$, and $\theta$
into the ans\" atze \eqref{n0} and \eqref{q-1} in order to get the
total density $\tilde{n}(\tilde{x})$, the condensate density $\tilde{n}_0(\tilde{x})$,
and the Bose-glass order parameter $\tilde{q}(\tilde{x})$.

\section{Results}
\label{IV}

\begin{figure}[!b]
\includegraphics[width=0.35\textwidth]{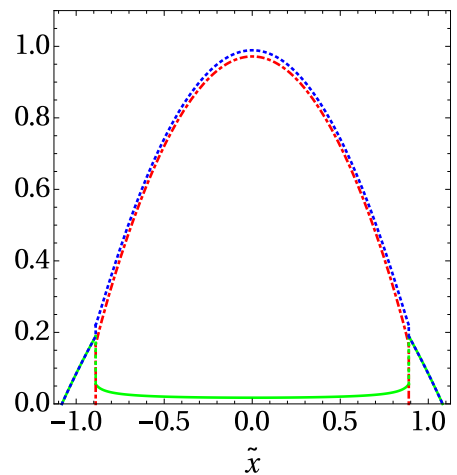}
\caption{TF results for the spatial distribution of the total particle
density $\tilde{n}(\tilde{x})$ (dotted, blue), condensate density
$\tilde{n}_0\left(\tilde{x}\right)$ (dotted-dashed, red), and Bose-glass
order parameter $\tilde{q}(\tilde{x})$ (solid green) for the disorder strength $\tilde{D}=0.016$.
\label{fig:2}}
\end{figure}

\begin{figure*}[!t]
\includegraphics[width=0.8\textwidth]{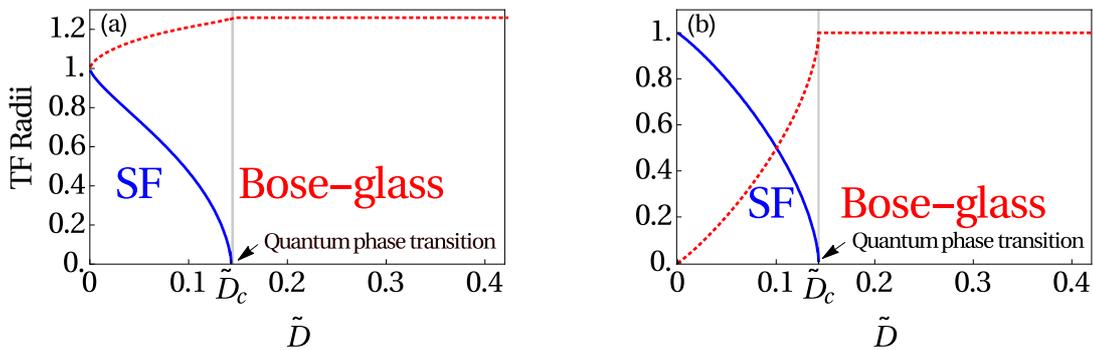}
\caption{TF results: (a) Cloud radius $\tilde{R}_{\rm TF2}$
(dotted, red) and condensate radius $\tilde{R}_{\rm TF1}$ (solid,
blue); (b) Fractional number of condensed particles $N_0/N$ (solid,
blue) and disconnected mini-condensates $Q/N$ (dotted, red)
as functions of the dimensionless disorder strength $\tilde{D}$. Both graphs reveal
a quantum phase transition from a superfluid (SF) to the Bose-glass phase.
\label{fig:3}}
\end{figure*}

The results presented in this section correspond to a dirty BEC with $N=10^6$ atoms of $^{87}$Rb,
with the $s$-wave scattering length $a=100\, a_0=5.29\,\mathrm{nm}$, where $a_0$ represents the Bohr radius. For the trap frequencies
we use experimentally realistic parameters: the longitudinal frequency is chosen to be $\Omega=2\pi\times50{\rm \,{Hz}}$, and the radial
one $\omega_{r}=2\pi\times179\,\mathrm{Hz}$.
For those parameters the longitudinal and the transversal oscillator lengths read $l=\sqrt{\frac{\hbar}{m\Omega}}=1.52\,\mu\mathrm{m}$ and
 $l_r=\sqrt{\frac{\hbar}{m\omega_r}}=806.04\,\mathrm{nm}$, respectively, while
the coherence length in the trap centre in the clean case turns out to be $\xi=45.6\,\mathrm{nm}$,
and the Thomas-Fermi radius reads $R_\mathrm{TF}=50.9\, \mu\mathrm{m}$.
Regarding the geometry of the system, we see that the transversal oscillator length
is much larger than the scattering length, $a\ll l_r$, but still smaller than the longitudinal oscillator length, $l_r<l$,
so we are indeed in the quasi one-dimensional regime \cite{J29,J30}.
If we estimate the value of the dimensionless quantity $\gamma_\mathrm{1D}=E_\mathrm{int}/E_\mathrm{kin}=Mg/\hbar^2 \overline{n}$ \cite{TF},
which compares the interaction and the kinetic energy of the system, where  the effective 1D interaction
strength is given by $g=2\hbar^2 Ma/l_r^2$ and $\overline{n}$ is defined in section~\ref{II}, we get $\gamma_\mathrm{1D}=2\times 10^{-7}\ll 1$.
This clearly shows that, for the chosen parameters, our system is in the weakly interacting regime,
and that we can describe it using the Hartree-Fock mean-field theory \cite{TF}.
Furthermore, if we calculate the value of the dimensionless quantity $\alpha_\mathrm{1D}=Mgl/\hbar^2=2al/l_r^2$, as defined in reference \cite{TF},
which relates the effective 1D interaction strength $g$ and the longitudinal trap frequency $\Omega$,
we obtain $\alpha_\mathrm{1D}=0.024\ll 1$. Since the condition $N\alpha_\mathrm{1D}\gg 1$ \cite{TF} is satisfied, we see that the system is close to the TF regime. 
This justifies our approach that, while using the TF approximation to obtain some analytical results and understand behaviour of the system in general,
full numerical treatment is still necessary in order to describe the properties of 1D dirty bosons.
In particular this will become obvious when we observe unphysical features of the TF approximation.

We first present results of the TF approximation and afterwards give in
parallel numerical and variational results and compare them.

\subsection{Thomas-Fermi results}

In section \ref{II} we have presented an analytical theory for the dirty boson problem in 1D.
Using the above specified parameter values, we now solve equation \eqref{PMM-1-1-1-1-1}
numerically. To this end we select from all its real solutions for $\tilde{n}_0(\tilde{x})$ the physical
one, i.e., the one with the smallest energy, and denote the region where it is non-trivial as a superfluid region.
The Bose-glass order parameter is here determined by equation \eqref{q1d-1-1}.
Then we combine the superfluid region solution with equation \eqref{bg},
describing the pure Bose-glass region, in which $\tilde{n}_0(\tilde{x})=0$ and $\tilde{n}(\tilde{x})=\tilde{q}(\tilde{x})$.
After that we fix the chemical potential $\tilde{\mu}$ using the
normalisation condition \eqref{18-1}. The resulting densities are
combined and plotted in figure~\ref{fig:2}.

The cloud radius $\tilde{R}_\mathrm{TF2}$ for the system is determined
by the condition that the total density vanishes, $\tilde{n}(\tilde{R}_\mathrm{TF2}=0$,
while the condensate radius $\tilde{R}_{\mathrm{TF}1}$ is maximal
value of the coordinate $\tilde{x}$ for which equation \eqref{PMM-1-1-1-1-1}
still has a solution, and separates the superfluid region ($|\tilde{x}|\leq\tilde{R}_{\mathrm{TF}1}$)
from the Bose-glass one ($\tilde{R}_\mathrm{TF1}<|\tilde{x}|\leq\tilde{R}_\mathrm{TF2}$).
This is illustrated in figure \ref{fig:2} for the dimensionless disorder strength $\tilde{D}=0.016$,
where we can see that the total density has a small jump at the condensate
radius, while the condensate density exhibits a jump to zero. The
Bose-glass order parameter $\tilde{q}(\tilde{x})$ has a double-bump
structure, exhibits a jump at the condensate radius, and also vanishes
at the cloud radius, by definition. In the Bose-glass region, the Bose-glass order
parameter and the total density coincide. The jump exhibited by all
three densities at the condensate radius is not a physical one, and
is an artefact of the applied TF approximation.

To study the influence of the disorder on the BEC properties, we plot the resulting
TF radii in figure~\ref{fig:3}(a) as a function of the disorder
strength $\tilde{D}$. We see that both cloud and condensate radius
coincide in the clean case, as expected. The condensate radius decreases with increasing
disorder strength and vanishes at the critical value $\tilde{D}_\mathrm{c}=0.143$,
which marks a quantum phase transition from the superfluid to the
Bose-glass phase. This corresponds to the value $\tilde{D}_\mathrm{c}=0.333$,
which was found in the non-perturbative approach of references \cite{Nattermann,Natterman2}
to be the critical disorder strength, where the Bose-glass phase becomes
energetically unstable and goes over into the superfluid phase. On
the other side, the cloud radius $\tilde{R}_\mathrm{TF2}$
increases with the disorder in the superfluid phase, but remains constant
in the Bose-glass phase at the value $\tilde{R}_\mathrm{TF2}=1.256$.
This means that beyond the critical disorder strength $\tilde{D}_\mathrm{c}$
the bosonic cloud is not extending anymore and has a maximal size.
The same conclusion can be deduced from figure~\ref{fig:3}(b),
where we depict the fractional number of condensed particles
$N_0/N=\frac{3}{4}\int_{-\tilde{R}_\mathrm{TF1}}^{\tilde{R}_\mathrm{TF1}}\tilde{n}_0\left(\tilde{x}\right)d\tilde{x}$.
Here $N_0/N$ equals to one in the clean case, i.e., all particles
are in the condensate. Afterwards it decreases as the disorder strength
$\tilde{D}$ increases, until it vanishes at $\tilde{D}_\mathrm{c}$,
marking the end of the superfluid phase and the beginning of the Bose-glass
phase. The fraction of the atoms in the disconnected local mini-condensates,
$Q/N=\frac{3}{4}\int_{-R_\mathrm{TF2}}^{R_\mathrm{TF2}}\tilde{q}\left(\tilde{x}\right)d\tilde{x}$,
behaves conversely. It increases with the increasing disorder until
reaching the maximal value of one at $\tilde{D}_\mathrm{c}$, then
it remains equal to one in the Bose-glass phase since all particles
are stuck in the local minima of the disorder potential.

\begin{figure*}[!ht]
\includegraphics[width=0.95\textwidth]{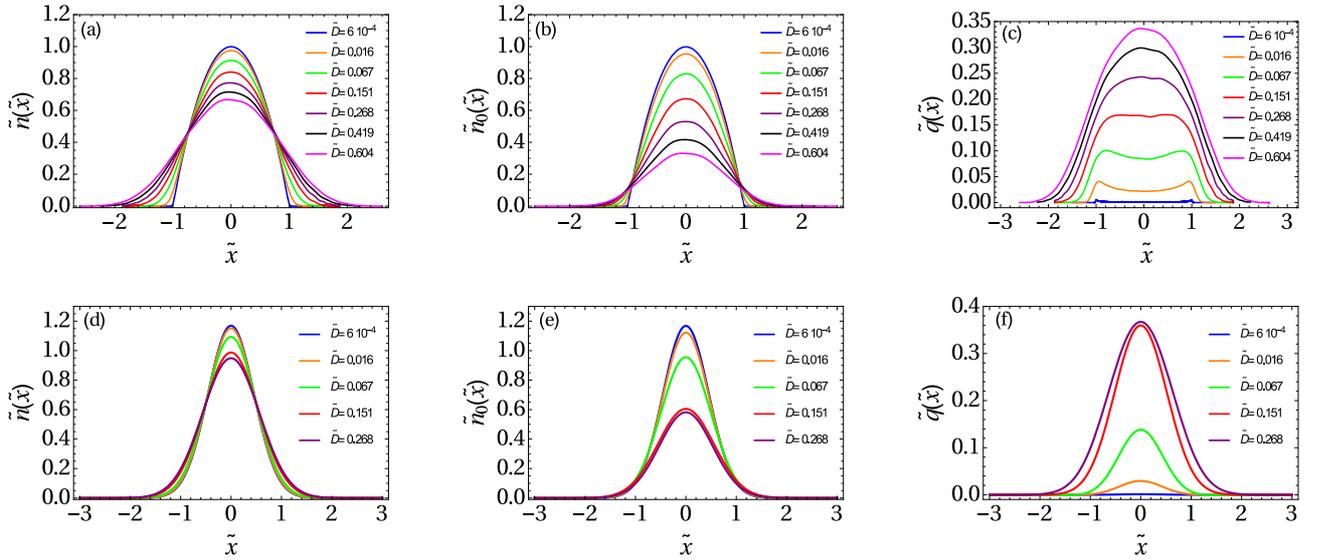}
\caption{Spatial distribution of numerically (top) and variationally (bottom) obtained: (a), (d) total particle density $\tilde{n}(\tilde{x})$;
(b), (e) condensate density $\tilde{n}_0(\tilde{x})$;
(c), (f) Bose-glass order parameter $\tilde{q}(\tilde{x})$,
for increasing disorder strengths $\tilde{D}$, from top to
bottom in the trap centre in (a), (b), (d), and (e), and from bottom
to top in (c) and (f).
\label{fig:4}}
\end{figure*}

\begin{figure*}[!ht]
\includegraphics[width=0.75\textwidth]{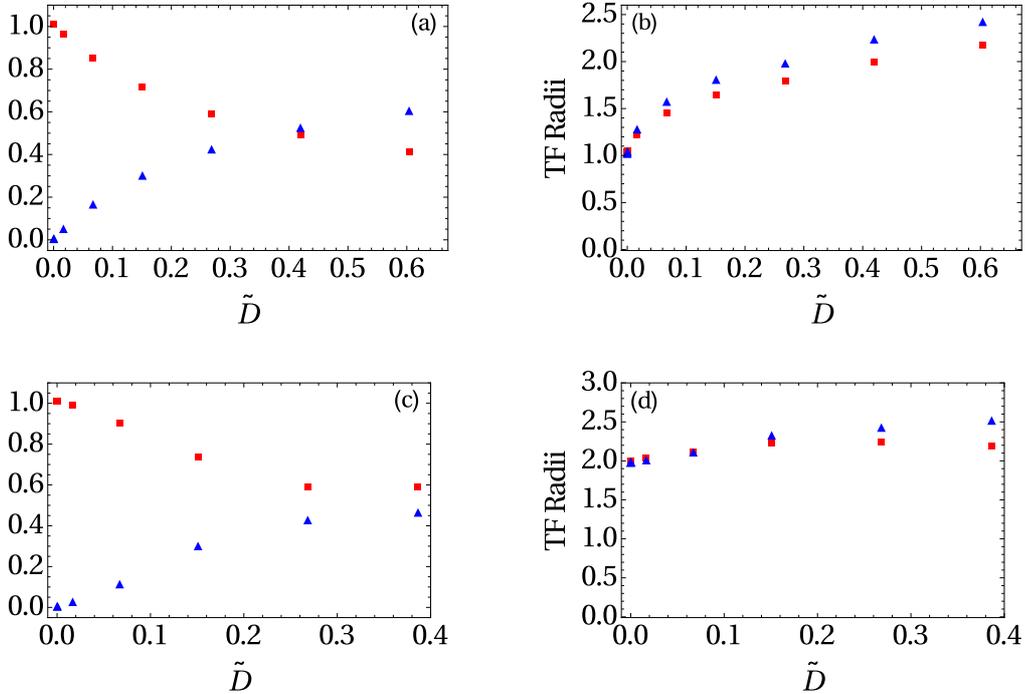}
\caption{(a) Numerical and (c) variational results for the
fractional number of condensed particles $N_0/N$ (red squares)
and fractional number of particles $Q/N$ in the disconnected local
mini-condensates (blue triangles) as functions of the
disorder strength $\tilde{D}$. (b) Numerical and (d) variational results for
the condensate radius $\tilde{R}_\mathrm{TF1}$ (red squares) and cloud
radius $\tilde{R}_\mathrm{TF2}$ (blue triangles) as functions of the
disorder strength $\tilde{D}$.
\label{fig:5}}
\end{figure*}

\subsection{Numerical and variational results}

Now we turn to numerical and variational results obtained using the
methods presented in section~\ref{III}. Figure~\ref{fig:4} presents in parallel numerically
and variationally obtained densities $\tilde{n}(\tilde{x})$, $\tilde{n}_0(\tilde{x})$,
and $\tilde{q}(\tilde{x})$ for various values of the disorder strength
$\tilde{D}$. The first notable difference compared to TF results
is that the condensate and the cloud radius are not clearly defined,
since the densities do not vanish at a well-defined point, but gradually
converge to zero at the respective borders. Therefore, we define the corresponding
radii, which, for simplicity, we again denote by $\tilde{R}_\mathrm{TF1}$
and $\tilde{R}_\mathrm{TF2}$, by the conditions $\tilde{n}_0(\tilde{R}_\mathrm{TF1})=\varepsilon$
and $\tilde{n}(\tilde{R}_\mathrm{TF2})=\varepsilon$, where $\varepsilon=10^{-4}$
represents a conveniently chosen small number. From figures~\ref{fig:4}(a) and \ref{fig:4}(d) we
see that the cloud radius increases with increasing disorder strength,
while the maximal density at the trap centre decreases. Figures~\ref{fig:4}(b)
and \ref{fig:4}(e) show the same type of behaviour for the condensate, and generally
reveal good agreement between the full numerical and variational results,
as for the total density. 

The numerically and variationally obtained values of the Bose-glass order parameter are also
plotted for different values of $\tilde{D}$ in figures~\ref{fig:4}(c) and \ref{fig:4}(f), respectively.
While the variational results have similar form as the total particle density and the condensate density,
due to the assumed functional dependence in equations (\ref{n0}) and (\ref{q-1}),
the numerical results show completely different behaviour.
In the weak disorder regime, the numerically calculated order parameter $\tilde{q}\left(\tilde{x}\right)$
reveals a double bump structure and is maximal at the border of the
condensate, while in the intermediate disorder regime it resembles a Gaussian-like
form. This redistribution takes place, according to figure~\ref{fig:4}(c),
at a disorder strength value between $\tilde{D}=0.151$ and $\tilde{D}=0.268$.
Thus, the main difference between the weak and the intermediate disorder
regime is that the local condensates concentrate at the border of
the condensate in the former case, but sit in the trap centre in the latter
case. Despite this marked difference, using either method yields that the width
as well as the maximum of the Bose-glass order parameter increase
with the disorder strength.

In order to obtain further information on the behaviour of the system, we plot in figures~\ref{fig:5}(a)
and \ref{fig:5}(c) the numerical and variational fractional
number of condensed particles $N_0/N$ and fractional number of particles in the disconnected mini-condensates $Q/N$,
as functions of the disorder strength, respectively. The condensed fraction $N_0/N$
decreases with the disorder strength and, conversely, $Q/N$ increases, meaning that more and more
particles are leaving the condensate with increasing $\tilde{D}$.
Unfortunately, the employed numerical algorithm breaks down for larger
values of the disorder strength, and one would have to use other approaches
in this case. Starting from the disorder strength value $\tilde{D}=0.393$,
the variational equations turn out to have only complex solutions,
so we cannot extract further information about our system for higher disorder
strengths using this approach. Therefore, we focus on the regimes of weak to moderate disorder.

Numerically and variationally calculated cloud radius $\tilde{R}_\mathrm{TF2}$ and the condensate radius
$\tilde{R}_\mathrm{TF1}$, as defined in this subsection, are plotted in figures~\ref{fig:5}(b)
and \ref{fig:5}(d) as functions of the disorder strength, respectively. Both radii are almost identical in the
weak disorder regime, and afterwards both increase linearly with the disorder
strength in the moderate disorder regime. Since the applied method breaks down for larger disorder strengths,
we cannot determine if a quantum phase transition exists, and this question remains still open.

\begin{figure*}[!ht]
\includegraphics[width=0.9\textwidth]{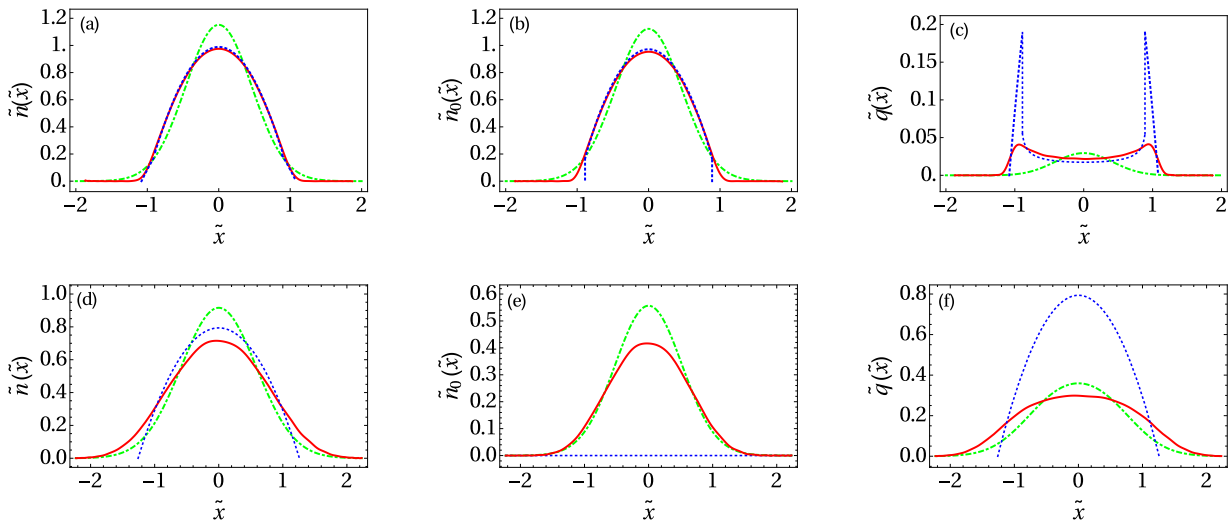}
\caption{A comparison of numerical (solid, red), variational (dotted-dashed, green), and TF-based analytical (dotted, blue) results
for $\tilde{D}=0.016$ (top) and $\tilde{D}=0.386$ (bottom) for:
(a), (d) total particle density $\tilde{n}(\tilde{x})$; (b), (e) condensate density $\tilde{n}_0(\tilde{x})$;
(c), (f) Bose-glass order parameter $\tilde{q}(\tilde{x})$.
\label{fig:6}}
\end{figure*}

\begin{figure*}[!ht]
\includegraphics[width=0.75\textwidth]{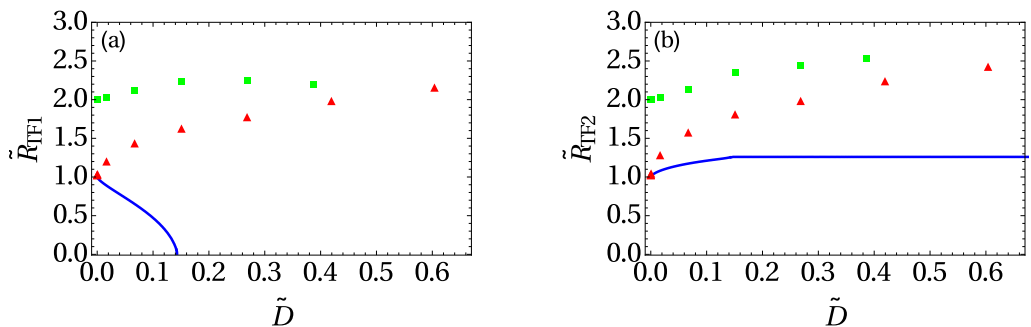}
\caption{A comparison of numerical (red triangles), variational (green squares), and analytical (solid blue line) results for (a) the condensate
radius $\tilde{R}_{\rm TF1}$ and (b) the cloud radius $\tilde{R}_{\rm TF2}$, as functions of the disorder strength $\tilde{D}$.
\label{fig:7}}
\end{figure*}

\subsection{Comparison}

Now we compare the physical quantities obtained via the three different
methods: the TF approximation, the numerical method, and the variational
method. For the small disorder strength $\tilde{D}=0.016$, the three
total densities $\tilde{n}(\tilde{x})$ in figure~\ref{fig:6}(a)
agree qualitatively well, but quantitatively the TF-approximated function $\tilde{n}(\tilde{x})$
is a better approximation for the numerical total density, especially
in the centre of the bosonic cloud, where the variational result does
not agree well with the numerical one. The same remark can be made
for the condensate density $\tilde{n}_0(\tilde{x})$ in figure~\ref{fig:6}(b).
For the Bose-glass order parameter $\tilde{q}(\tilde{x})$ in figure~\ref{fig:7}(c)
the double-bump structure, which exists in both numerical and TF-approximated
results, is missing in the variational result, which has just a bell
form, as assumed by the variational ansatz. This makes again the TF-approximated Bose-glass
order parameter $\tilde{q}(\tilde{x})$ a better approximation for
the numerical one.

For the moderate disorder strength $\tilde{D}=0.386$,
the TF-approximated total density $\tilde{n}(\tilde{x})$ in figure~\ref{fig:6}(d)
is also a better approximation for the numerical one in the centre
of the bosonic cloud, while at the trap borders the variational approximation wins.
According to the TF results shown in figure~\ref{fig:3}, at the disorder strength
value $\tilde{D}=0.386$ we are already in the Bose-glass phase, thus, the
TF-approximated condensate density $\tilde{n}_0(\tilde{x})$ in
figure~\ref{fig:6}(e) vanishes. This is not the case for
both the numerical and the variational condensate densities, which
are compatible and match quite well at trap borders. The variational
Bose-glass order parameter $\tilde{q}(\tilde{x})$ in figure~\ref{fig:6}(f)
also agrees well with the numerical one and both have the same bell
shape, while the TF-approximated Bose-glass order parameter has a
significant deviation, which is expected since the TF approximation
breaks down in the moderate disorder regime.

In figures~\ref{fig:7}(a) and \ref{fig:7}(b) we see that the variational and the numerical condensate radius $\tilde{R}_\mathrm{TF1}$
and cloud radius $\tilde{R}_\mathrm{TF2}$ have the same behaviour, namely both increase with the disorder strength.
This is in stark contrast to the TF result, where the condensate radius is found to decrease with $\tilde{D}$,
evenatually leading to a quantum phase transition at $\tilde{D}_\mathrm{c}=0.143$.
Such a quantum phase transition is predicted only in TF approximation, which is known to fail for moderate disorder strengths.

The question that still remains to be answered
concerns the possible existence of the quantum phase transition from
the superfluid to the Bose-glass phase for large disorder strengths.
According to reference \cite{Natterman2},
the disorder has to energetically overcome the interaction in order
to yield such a quantum phase transition. However, numerical
and variational results displayed in figure~\ref{fig:7}(a)
suggest that this is not the case neither in the weak nor in the moderate
disorder regime. One would have to investigate much stronger disorder
regime, employing a different set of methods, in order to be able to detect a possible quantum phase transition,
which is beyond the scope of this paper.

\section{Conclusions}

From the discussion in the previous section, we conclude that the
TF approximation yields good results for the quasi-one-dimensional
dirty bosons in the weak disorder regime, which agree well with the
numerical ones, especially in the centre of the bosonic cloud, where
the kinetic energy can be neglected. However, this approximation breaks
down in the moderate disorder regime, and is unable to describe
the dirty BEC system properly. The origin of this failure is the fact
that the condition \eqref{TF} is not fulfilled in the moderate
disorder regime. The coherence length becomes significantly larger
as we increase the disorder strength, especially at the border of
the bosonic cloud, and when it becomes of the order of the Thomas-Fermi
radius, the TF approximation breaks down. Furthermore, quantum fluctuations
are more prominent in lower dimensions, which also restricts the
validity range of the TF approximation. On the other side, the variational
method with the ans\" atze \eqref{n0}--\eqref{Q0-2} turns out to be
a good approximation to describe the dirty BEC system in the moderate
disorder regime and works there better than in the weak disorder regime, especially
at the cloud border, where the Bose-glass region is located.
This is due to the fact that a stronger disorder reduces significantly
the repulsive interaction between the particles and approaches the
case of non-interacting bosons, where the densities are Gaussian-like,
as in our variational ans\" atze. Although the variational method breaks down for larger disorder strengths, it still
provides results in an important range of disorder strengths. The combination
of applying the TF approximation for the weak disorder together
with the variational method for the moderate disorder covers
a significant range of disorder strengths. With this we can analytically describe
the redistribution of the local disconnected mini-condensates from the edge of the atomic
cloud to the trap centre for increasing disorder strengths, as obtained
from detailed numerical simulations. We expect that all these results
are useful for a quantitative analysis of experiments for dirty bosons
in quasi-one-dimensional harmonic traps. The problem of the large
disorder strengths still persists with the current approach and remains
to be addressed by other methods.

\begin{acknowledgments}
The authors gratefully acknowledge financial support from German Academic
and Exchange Service (DAAD), the Ministry of Education, Science, and
Technological Development of the Republic of Serbia under projects
ON171017, NAI-DBEC, and IBEC, and German Research Foundation (DFG)
via the Collaborative Research Center SFB/TR49 ``Condensed Matter
Systems with Variable Many-Body Interactions''. 
\end{acknowledgments}

Numerical simulations were run on the PARADOX supercomputing facility
at the Scientific Computing Laboratory of the Institute of Physics
Belgrade.

\appendix

\section{Free energy}
\label{F}

Here we briefly summarise the main result of reference \cite{Paper2},
which relies on deriving a Hartree-Fock approximation for the free
energy of one-dimensional harmonically trapped dirty bosons.
The starting point is the functional integral for the grand-canonical
partition function

\begin{equation}
\mathcal{Z}=\mathcal{\oint}\mathcal{D}\psi^{\ast}{\displaystyle \oint}\mathcal{D}\psi e^{-\mathcal{A}\left[\psi^{\ast},\psi\right]/\hbar},\label{eq:Z}
\end{equation}
where the integration is performed over all Bose fields $\psi^{*}(x,\tau),\psi(x,\tau)$,
which are periodic in imaginary time $\tau$, i.e.,
$\psi(x,\tau)=\psi(x,\tau+\hbar\beta)$. The
Euclidean action is given in standard notation by 
\begin{widetext}
\begin{eqnarray}
\mathcal{A}[\psi^*,\psi] & =&\int_0^{\hbar\beta}d\tau\int dx\Bigg[ \psi^*(x,\tau)\Big\{\hbar\frac{\partial}{\partial\tau}-\frac{\hbar^2}{2M}\Delta
+V(x)+U(x)-\mu\Big\}\psi(x,\tau)\nonumber\\
&&+\frac{1}{2}\int dx'\psi^*(x,\tau)\psi(x,\tau)V^{({\rm int})}(x-x')\psi^*(x',\tau)\psi(x',\tau)\Bigg]\, ,\label{eq:A}
\end{eqnarray}
\end{widetext}
where $V(x)=M\Omega^2x^2/2$ denotes the harmonic trap with the
trap frequency $\Omega$, $M$ the particle mass, $\mu$ the chemical
potential, and $V^{({\rm int})}(x-x')=g\delta(x-x')$ the contact
interaction potential. The interaction coupling strength $g=2a\hbar\omega_{r}$
depends on the s-wave scattering length $a$, which has to be positive
in order to obtain a stable BEC, and the transversal trap frequency
$\omega_{r}$. Note that the latter has to be large enough, i.e.,
$\omega_{r}\gg\Omega$, in order to ensure a quasi-one-dimensional
setup \cite{J29,J30}.

We assume for the disorder potential $U(x)$ that it is homogeneous
after performing the disorder ensemble average (denoted by $\overline{\bullet}$)
over all possible realisations. Thus, the expectation value of the
disorder potential can be set to vanish without loss of generality, as defined by equation \eqref{eq:19}.
The disorder correlation function is given by equation \eqref{1}, where 
we assume $D(x-x')=D\,\delta(x-x')$, and $D$ denotes the disorder strength.

Within the Hartree-Fock mean-field approximation with the replica
method, reference \cite{Paper2} obtains self-consistency equations, which
determine the particle density $n(x)$ as well as the order parameter
of the superfluid $n_0(x)$, which represents the condensate density,
and the order parameter of the Bose-glass phase $q(x)$, which stands
for the density of the particles being condensed in the respective
minima of the disorder potential.

More precisely, the two order parameters $n_0(x)$ and $q(x)$ of
our mean-field theory at $T=0$ are defined by following the notion
of spin-glass theory \cite{Intro-101,Intro-35,Hertz}. On the one hand, the off-diagonal
long-range limit of the two-point correlation function defines the condensate density \cite{Leggett},
\begin{equation}
\label{A5}
\lim_{|x-x'|\to\infty}\overline{\langle\psi(x,\tau)\psi^{*}(x',\tau)\rangle}=\sqrt{n_0(x) n_0(x')}\,.
\end{equation}
On the other hand, the Bose-glass order parameter $q(x)$ was introduced in
reference \cite{Intro-90} in close analogy to the Edward-Anderson order
parameter of spin-glasses \cite{19-1} by the off-diagonal long-range limit of the four-point correlation function,
\begin{widetext}
\begin{equation}
\lim_{|x-x'|\to\infty} \overline{\left|\langle\psi(x,\tau)\psi^{*}(x',\tau)\rangle\right|^2}=
[n_0(x)+q(x)][n_0(x')+q(x')]\, .
\label{A6}
\end{equation}
\end{widetext}

Note that the validity of equations \eqref{A5} and \eqref{A6} within the Hartree-Fock approximation was analysed in detail in reference \cite{Paper2}.
There it is also shown that two- and four-point correlations decay exponentially on a length scale which can be physically interpreted as the localisation length
named after Larkin \cite{Nattermann,Larkin}. A similar exponential decay of the one-body density matrix in the presence of disorder was found in references
\cite{Intro-11,Ref2-6,Ref2-7}, although in the quasi-condensed phase a power-lay decay is expected as in the spatially uniform gas \cite{Ref2-4,Ref2-5}.

In the one-dimensional case and at $T=0$, the mean-field Hartree-Fock
theory with the help of the replica method leads to the free energy \cite{Paper2}:
\begin{widetext}
\begin{eqnarray}
\mathcal{F} & = &\int dx\Bigg[ -g\left[q(x)+n_0(x)\right]^2-\sqrt{n_0(x)}
\left\{\mu+\frac{\hbar^2}{2M}\frac{\partial^2}{\partial x^2}-2g\left[q(x)+n_0(x)\right]-\frac{1}{2}M\varOmega^2 x^2
+\frac{D}{\hbar}Q_0(x)\right\}\sqrt{n_0(x)}\nonumber\\
&& -\frac{g}{2}n_0(x)^2+\frac{D}{\hbar}Q_0(x)[q(x)+n_0(x)]-\frac{D}{\hbar}\sqrt{\frac{M}{2}}
 \frac{q(x)+n_0(x)}{\sqrt{-\mu+2g\left[q(x)+n_0(x)\right]+\frac{1}{2}M\varOmega^2 x^2-\frac{D}{\hbar}Q_0(x)}}\Bigg]\, .\label{F1}
\end{eqnarray}
\end{widetext}
Here $Q_0(x)$ represents an auxiliary function within the Hartree-Fock
theory. From the thermodynamic relation $-\frac{\partial\mathcal{F}}{\partial\mu}=N$
we obtain 
\begin{equation}
\int_{-\infty}^\infty n(x)dx=N\, ,\label{3.5}
\end{equation}
which defines the particle density $n(x)$. Extremizing the free energy \eqref{F1} with respect to $n_0(x)$, $q(x)$, and $Q_0(x)$ yields,
together with equation \eqref{3.5}, the self-consistency equations \eqref{q1d}--\eqref{Q3}.

\section{Homogeneous case}
\label{homogeneous}

The simplest case to discuss for dirty bosons is the homogeneous one, where $V(x)=0$.
Since all densities are spatially constant in the homogeneous case,
we drop in this section the $x$ dependency of all densities. With
this, equations \eqref{q1d}--\eqref{Q3} reduce, after eliminating
$Q_0$, to:
\begin{eqnarray}
q&=&\frac{D}{\hbar M}\,\frac{n_0}{\left(\frac{2gn_0}{M}\right)^{\frac{3}{2}}-\frac{D}{\hbar M}}\, ,\label{q1d-1-2}\\
gn_0&=&-\mu+2gn-\frac{D}{\hbar}\sqrt{\frac{M}{2gn_0}}\, ,\label{PMM-1-1-2}\\
n&=&q+n_0\, .\label{MAT1-1-2}
\end{eqnarray}
From equations \eqref{q1d-1-2} and \eqref{MAT1-1-2} we get an algebraic
fifth-order equation for the condensate fraction $n_0/n$: 
\begin{equation}
\left(\frac{n_0}{n}\right)^{5/2}-\left(\frac{n_0}{n}\right)^{3/2}+\overline{D}=0\, ,\label{16}
\end{equation}
where $\overline{D}=\frac{\xi^{3}}{\mathcal{L}^3}$ denotes the
dimensionless disorder strength, $\xi=\frac{\hbar}{\sqrt{2Mgn}}$
the coherence length, and $\mathcal{L}=\left(\frac{\hbar^4}{M^2D}\right)^{1/3}$
the Larkin length \cite{Larkin,Nattermann}.

\begin{figure}[!b]
\includegraphics[width=0.4\textwidth]{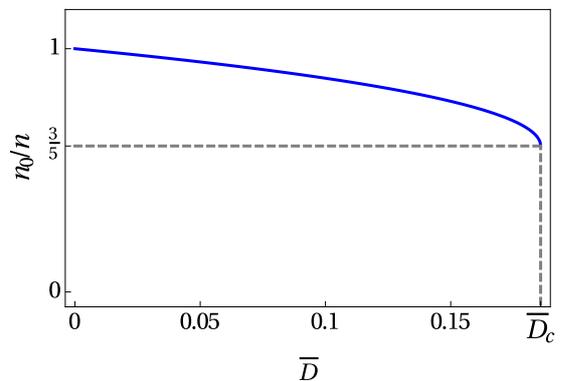}
\caption{Condensate fraction $n_0/n$ as function
of dimensionless disorder strength $\overline{D}$ as a solution
of equation \eqref{16}.
\label{fig:8}}
\end{figure}

Figure~\ref{fig:8} shows how condensate fraction $n_0/n$ decreases with increasing the
disorder strength $\overline{D}$ according to equation \eqref{16}.
Thus, the Hartree-Fock mean-field theory predicts a first-order quantum phase transition
from the superfluid phase to the Bose-glass phase at the critical
value $\overline{D}_{\rm c}=\frac{6}{25}\sqrt{\frac{3}{5}}\simeq 0.185$.
This corresponds to the value $\overline{D}_{\rm c}=1$, that
was found in the non-perturbative approach of references \cite{Nattermann,Natterman2},
which investigate at which disorder strength the Bose-glass phase
becomes energetically unstable and goes over into the superfluid phase.
Therefore, we expect that a quantum phase transition will also appear
in the trapped case within the Thomas-Fermi approximation. 

Now we check whether our results are compatible with the Huang-Meng
theory \cite{HM-1,HM-3,HM-4,HM-5,HM-2,HM-8,HM-6,HM-9,HM-10,HM-7,Ref2-3,Ref2-7},
where the Bose-glass order parameter of a homogeneous dilute Bose
gas at zero temperature in case of weak disorder regime is deduced
within the seminal Bogoliubov theory. The Bose-glass order parameter
in one dimension is according to the Huang-Meng theory proportional
to the disorder strength, which yields in dimensionless form 
\begin{equation}
\frac{q_{\rm HM}}{n}=\frac{\overline{D}}{2^{3/2}}\, .
\label{3.23}
\end{equation}

In our Hartree-Fock mean-field theory the Bose-glass order parameter
in case of weak disorder strength turns out to be:
\begin{equation}
\frac{q_{w}}{n}=\overline{D}\, .
\label{19-1}
\end{equation}

Thus, from equation \eqref{19-1} we conclude that our
result agrees qualitatively with the Huang-Meng theory. But quantitatively
the comparison of equations \eqref{3.23} and \eqref{19-1} reveals that
a factor of $2^{3/2}$ is missing in our result \eqref{19-1}. This
is due to the fact that the Hartree-Fock theory does not contain the
Bogoliubov channel, which is included in the Huang-Meng theory.

\end{document}